\documentclass[a4paper]{jpconf}
\usepackage{graphicx}
\usepackage{epsfig}
\usepackage{amsmath}
\usepackage{amsfonts}

\def\ba{\begin{eqnarray}}
\def\ea{\end{eqnarray}}

\def\gtorder{\mathrel{\raise.3ex\hbox{$>$}\mkern-14mu
\lower0.6ex\hbox{$\sim$}}}
\def\ltorder{\mathrel{\raise.3ex\hbox{$<$}\mkern-14mu
\lower0.6ex\hbox{$\sim$}}}

\begin{document}

\rightline{DF/IST-1.2009}
\rightline{LPT/09-06}

\title{Modelling non-Gaussianity from foreground contaminants}
\author{
C.S.~Carvalho\footnote[1]{Present address}$^{,2,3}$}
\address{
$^1$Astrophysics and Cosmology Research Unit, 
School of Mathematical Sciences, 
University of KwaZulu-Natal, 
Durban, 4041, South Africa} 
\address{
$^2$Laboratoire de Physique Th\'eorique, B\^atiment 210, 
Universit\'e Paris-Sud, 91405 Orsay Cedex, France}
\address{
$^3$Instituto Superior T\'ecnico, Departamento de F\'isica, 
Avenida Rovisco Pais, 1049-001 Lisbon, Portugal}

\ead{Carla-Sofia.Carvalho@th.u-psud.fr}

\begin{abstract}
We introduce a general class of models for charaterizing the
non-Gaussian properties of foreground contaminants in the cosmic
microwave background with view towards the removal of the 
non-primordial non-Gaussian signal from the primordial
one. 
This is important not only for treating temperature maps but also for
characterizing the nature and origin of the primordial cosmological
perturbations and thus establishing a theory of the early universe.
\footnote[4]{This manuscript is based on a talk delivered by the
  author at the NEB XIII, Thessaloniki, June 2008} 
\end{abstract}

\section{Introduction}

Cosmology is presently a very active field because of the large number of
observations that are becoming available and that will allow us to characterize
with great precision the nature and physical origin of the primordial
cosmological perturbations. A key question is whether these primordial
perturbations were Gaussian.

The currently favoured best-fit cosmological models, as supported by
the recent Wilkinson Microwave Anisotropy Probe (WMAP) 
data, 
are in agreement with the primordial cosmological perturbations being
Gaussian to great accuracy \cite{wmap5:komatsu}.
With the WMAP data there has been a claimed detection of primordial
non-Gaussianity that is currently disputed \cite{wandelt08}.
However, even the conservative official analysis hints at something,
the value of $f_{NL}$ lying within $1.9\sigma$ from the null value.


Either a detection or a more stringent constraint of
 $f_{NL}$
would have profound implications for our understanding
of the physics of the early Universe.
In particular, it would be an observational discriminant among competing
models for the generation of primordial fluctuations.
An $f_{NL}\not=0$ would be a very
interesting challenge for cosmology since it would require an
extensive revision of the standard cosmological model.
Present observations
have ruled out a great number of cosmological models already.
All currently favoured models would be ruled out by such a detection.
A non-detection of $f_{NL}$ with
an improved constraint on the allowed values
would strongly restrict the fluctuation generation mechanisms.

Due to the great improvement in sensitivity of the European Space
Agency's {\it Planck} satellite over WMAP,
understanding the nature and in particular non-Gaussian properties of
foregrounds will be indispensable for the {\it Planck} $f_{NL}$
analysis. For WMAP, 
foregrounds were not an issue because no signal was detected. Currently
estimators exist \cite{babich05,creminelli06} that are proven optimal
in the absence of foreground contamination and much work has been
devoted to their efficient implementation \cite{wmap5:gold}.
However, to date there has been no comprehensive study of the
impact of foregrounds.
Foreground sources distributed in an anisotropic way are susceptible
of being misidentified as primordial non-Gaussianity.
In order to extract from the data an accurate and significant measure
of $f_{NL},$  new estimators are needed that divide the data and are
designed to mask possible contaminants. First, however,
it is required to characterize the statistical
properties of these spurious sources of non-Gaussianity, which
requires understanding the non-linear properties of foregrounds.

This contribution reports on work in progress
where we formulate a general class of models
for characterizing the non-Gaussian properties of
foreground contaminants in the cosmic microwave background (CMB)
\cite{bucher}.

\section{Measuring non-Gaussianity}

In linear perturbation theory, the temperature anisotropy of the CMB
is derived from metric perturbations by the relation 
$
{\Delta T/T}
\propto g_{T}\Delta\Phi,$
where the radiation transfer function $g_T$ depends on the scale
considered.\footnote{
For temperature fuctuations on super-horizon scales 
the Sachs--Wolfe efect 
dominates and $g_T$ is a function of the parameter $\omega$ of the
equation of state. At the decoupling epoch it
reduces to $g_{T}= -1/3$ in the case of adiabatic fluctuations. 
On sub-horizon scales, however, $g_T$ oscillates and we need 
to solve the Boltzmann photon transport equations coupled to the
Einstein equations for $g_T,$ which is found to have as parameter the
decoupling time.} 
Linear perturbation theory is a valid approximation because  
the physics of the pertubations generated during inflation is
extremely weakly coupled.

In the simplest models of inflation, metric perturbations are produced
by quantum fluctuations of the inflaton field. 
If the physics that governs the evolution of the metric perturbations
is linear, since quantum fluctuations are Gaussian then so will
temperature fluctuations be Gaussian. 
However, 
non-linear coupling between long and short wavelength fluctuations of
the inflaton can generate weak non-Gaussianities. These will propagate
into the metric perturbations and consequently into the temperature
anisotropy. 
The primordial metric perturbations $\Phi$ can be characterized
approximately by a non-linear coupling parameter, $f_{NL},$ such that
\cite{komatsu00} 
\ba
\Phi({\bf x}) 
=\Phi_{L}({\bf x}) 
+f_{NL}\left[ \Phi_{L}^2({\bf x}) - \left< \Phi_{L}^2({\bf x})\right>\right],
\ea
where $\Phi_{L}({\bf x})$ are Gaussian linear perturbations with 
$\left< \Phi_{L}({\bf x})\right> =0.$
In many models (e.g. single-field slow roll inflation 
\cite{inflation_ena_a, inflation_ena_b}), 
the non-linear corrections are too small to be detected.
Other models (e.g. multiple field inflation 
\cite{inflation_polla_a, inflation_polla_b})  
can generate stronger and potentially detectable
non-linearity. 
A Gaussian distribution is characterized by vanishing odd-order moments
and even-order moments defined only by the variance. Non-Gaussianity can
thus be measured by correlations among an odd number of points. 
Here we shall use 
the angular three-point correlation function or
bispectrum as our statistical tool since it is sensitive to weakly non-Gaussian
fluctuations.

Expanding the temperature anisotropy in momentum space
\ba
{\Delta T(\Omega)\over T}
=\sum_{l,m} a_{lm}Y_{lm}(\Omega),
\ea
where the harmonic coefficients are
\ba
a_{lm}
=4\pi(-i)^l\int {{d^3k}\over {(2\pi)^2}}
 \Phi({\bf k})g_{T_l}({\bf k})Y_{lm}^{\ast}(\Omega),
\ea
the bispectrum is defined as
\ba
B_{l_1l_2l_3}^{m_1m_2m_3}
\equiv \left< a_{l_1m_1}a_{l_2m_2}a_{l_3m_3} \right>
=G_{l_1l_2l_3}^{m_1m_2m_3}b_{l_1l_2l_3}.  
\ea
Here
\ba
G_{l_1l_2l_3}^{m_1m_2m_3}
=\int d\Omega Y_{l_1m_1}(\Omega)Y_{l_2m_2}(\Omega)Y_{l_3m_3}(\Omega)
\ea
is the Gaunt integral and
\ba
b_{l_1l_2l_3}
&=&\left({2\over \pi}\right)^3\int dx~dk_1~dk_2~dk_3~(xk_1k_2k_3)^2
  B_{\Phi}(k_1,k_2,k_3)\cr
&&\times j_{l_1}(k_1x)j_{l_2}(k_2x)j_{l_3}(k_3x)~
   g_{T_{l_1}}(k_1)g_{T_{l_2}}(k_2)g_{T_{l_3}}(k_3)
\ea
is the reduced bispectrum defined in terms of the radiation transfer
functions $g_{T_{l}},$ the Bessel functions of fractional order $j_{l}$ 
and the primordial bispectrum $B_{\Phi}(k_1,k_2,k_3).$
In the flat-sky approximation, which amounts to $(l,m) \to \boldsymbol{\ell}$,
where $\boldsymbol{\ell}$ is a two-dimensional wave vector in the sky,
the bispectrum reduces to 
\ba
\left<
a(\boldsymbol{\ell}_1)a(\boldsymbol{\ell}_2)a(\boldsymbol{\ell}_3)\right>
=(2\pi)^2\delta^2(\boldsymbol{\ell}_1+\boldsymbol{\ell}_2+\boldsymbol{\ell}_3)
B(\boldsymbol{\ell}_1,\boldsymbol{\ell}_2,\boldsymbol{\ell}_3)
\ea
which is non-vanishing only for closed triangle configurations in momentum
space. Since in this approximation 
$G_{l_1l_2l_3}^{m_1m_2m_3}\to 
(2\pi)^2\delta^2(\boldsymbol{\ell}_1+\boldsymbol{\ell}_2+\boldsymbol{\ell}_3),$
then $b_{l_1l_2l_3}\to 
B(\boldsymbol{\ell}_1,\boldsymbol{\ell}_2,\boldsymbol{\ell}_3)$ 
\cite{komatsu04}.

The signal-to-noise squared for the bispetrum gives us an indication
of the dominant triangle configuration to the non-Gaussian
signal.
We assume a scale invariant power spectrum 
$P(\ell)\propto \ell^{-2}$ in the flat sky approximation and 
$B(\ell_1, \ell_2, \ell_3)
=f_{NL}\left[
P(\ell_1)P(\ell_2) +P(\ell_2)P(\ell_3) +P(\ell_3)P(\ell_1)\right]$
as a result of the local non-Gaussianity ansatz.
Then 
\ba
\left( {S\over N}\right)^2 
&\propto&
 \int d^2\ell_1 
  \int d^2\ell_2 
   \int d^2\ell_3 
    \delta^2(\boldsymbol{\ell}_1+\boldsymbol{\ell}_2+\boldsymbol{\ell}_3)
{{B^2(\ell_1,\ell_2,\ell_3)}
 \over {P(\ell_1) P(\ell_2) P(\ell_3)}}\cr
&\propto& f_{NL}^2
 \int
 d^2\ell_1
  \int
  d^2\ell_2
   \int
   d^2\ell_3 
    \delta^2(\boldsymbol{\ell}_1+\boldsymbol{\ell}_2+\boldsymbol{\ell}_3)
{{\left( 
{1\over {\ell_1^2\ell_2^2}} 
+{1\over {\ell_2^2\ell_3^2}} 
+{1\over {\ell_3^1\ell_1^2}}\right)^2}
 \over {1\over {\ell_1^2\ell_2^2\ell_3^2}}}\cr
&\propto& f_{NL}^2 \int{d\ell_{<}\over \ell_{<}}\int d^2\ell_{>}.
\ea
For squeezed triangles, where the length of the two large sides is of
order $\ell_{max}$ and that of the smaller side $\ell_{min},$ we find that
\ba
\left({S\over N}\right)^2
\propto& f_{NL}^2 \int{d\ell_{<}\over \ell_{<}}\int d^2\ell_{>}
\propto  f_{NL}^2 
 \ln\left[{\ell_{max}\over \ell_{min}}\right]
  \ell_{max}^2.
 \ea
For equilateral triangles with the length of the sides of
order $\ell_{max}$ 
\ba
\left({S\over N}\right)^2 
\propto f_{NL}^2 \ell_{max}^2.
\ea
Comparing the two results we observe that 
the dominant contribution to the signal comes from the squeezed
triangles, which introduce the logarithmic boost factor and thus
couple the small-scale with the large-scale anisotropies. 
Should only equilateral triangles be considered, then the
description of the signal would be equivalent to the separate
description from different patches in the sky, which would mean losing
most of the signal. 
The squeezed triangle configurations correspond to the local 
case where the non-Gaussianity was created primarily on super-horizon
scales, as opposed to the equilateral case where the non-Gaussianity
is primarily created at horizon-crossing. 
Qualitatively an $f_{NL}\not= 0$ can be understood as a modulation
of the small-scale power (on scales near the resolution of the map) by
the large-scale signal. 
Exploring further this idea, we can construct a cubic estimator of
$f_{NL}$ by filtering the
signal into two contributions, namely $T_{\text{low $\ell$}}$ and 
$T_{\text{high $\ell$}},$ and considering the integral
\ba
E=\int d\Omega~T_{ \text{low $\ell$}}T_{ \text{high $\ell$}}^2.
\ea
This is but a caricature of the 
optimal estimator given in Ref.~\cite{spergel03}
with the ideal weighting, which here serves illustration purposes. 
The large-scale variation in $T_{ \text{high $\ell$}}^2$  reflects
spatial dependence in the power spectrum $P(\ell,\Omega)$ where $\ell$ is large
and the variation in $\Omega$ is slow. In a Gaussian
model, such perceived spatial dependence is simply noise from the
cosmic variance and thus fortuituous. In a model with $f_{NL}\not=0,$
however, it is correlated with the large-scale temperature
anisotropy. 

Astronomical microwave sources as well as instrumental effects can
produce spurious non-Gaussian signals 
which would be measured together
with the primordial signal by the bispetrum and thus susceptible 
of being confused for primordial non-Gaussianity.
Examples of such sources are \cite{wmap1:bennet}: 
the synchroton emmision, which results from the acceleration of cosmic ray
electrons in magnetic fields; 
the free-free emission, which is produced by electron-ion scattering
and is thus correlated with Hydrogen emission lines;
and thermal galactic dust emission. 

The separation of the CMB and foreground signal components has relied on
their differing spectral and spacial distribution.
However, correlations at different scales of the power of
non-primordial foreground signals could not be separated
from the CMB by the canonical methods.
Here we aim to show that a detectable non-Gaussian signal could be mimicked
by such correlations. 

\section{Faking non-Gaussianity}

Since the $a_{lm}$'s are random quantities for the same sky
realization, we must devise a way of generating maps of
the sky with built-in non-linear foregrounds. 
We use motifs of a fixed shape and distributed in a probabilistic
manner in the sky to model foreground sources. This resembles the
galaxy counting method for the study of the large-scale structure
formation \cite{peebles}.

Foreground realizations are generated in two steps. 
First a point $\boldsymbol{\xi}$ for the centre of a motif is
randomly selected according to a probability density function
$Q(\boldsymbol{\xi})$ in the simplest case in an uncorrelated
Poissonian manner. 
Then to each point $\boldsymbol{\theta}$ we assign a conditional
probability $P(\boldsymbol{\theta },\boldsymbol{\xi })$  of finding
the point $\boldsymbol{\theta }$ for a motif centered at
$\boldsymbol{\xi }.$  This probability is independent of how many
points are already inside the motif. 
We assign the value one to the motif interiors and zero to the regions
between motifs. Assuming that the foregrounds are optically thin,
i.e. that they do not mask the effects of each other, we take
overlapping motifs to add.  
We also need to properly normalize these quantities so as to relate to the
foreground maps. 
Measuring the mean number density of motifs,
the probability density $Q(\boldsymbol{\xi})$ is such that when
integrated over the entire sky it yields the average number of motifs
$\left<N\right>$  
\ba
\int d^2\xi~Q(\boldsymbol{\xi})=
\left<\text{number of motifs}\right>.
\ea
The conditional probability $P(\boldsymbol{\theta},\boldsymbol{\xi})$ 
yields the temperature at a point $\boldsymbol{\theta }$ given that a
motif is centred at $\boldsymbol{\xi }.$
Thus correlations among the $\xi$'s as encoded in $Q$ will model the
large-scale structure, 
while correlations among the $\theta$'s in the same motif as encoded
in $P$ will model the small-scale structure.

We calculate the two and three-point correlation functions
of the statistical map in a perturbative manner using Feynman
diagrams. These correlation functions are interpreted
respectively as the power spectrum and the bispetrum by assigning a
temperature to the motifs. This temperature is contrained: from below
by the minimum temperature that a foreground distribution must have to
account for the remaining signal after removal of the astrophysical
sources described previously; 
and from above by the maximum value of the power spectrum that a
spurious source could have so as not to overshadow the background
radiation.

We define the two-point correlation function of the final map
${\mathcal P}(\boldsymbol{\theta }_1, \boldsymbol{\theta }_2)$ 
by the sum over the two contributions to the probability that
$\boldsymbol{\theta }_1$ and 
$\boldsymbol{\theta }_2$ are within a motif
\ba
{\mathcal P}(\boldsymbol{\theta }_1, \boldsymbol{\theta }_2)
={\mathfrak p}_{1,1}(\boldsymbol{\theta }_1, \boldsymbol{\theta }_2)
+{\mathfrak p}_{2}(\boldsymbol{\theta }_1, \boldsymbol{\theta }_2),
\ea
where
\ba
{\mathfrak p}_{1,1}&=&
\frac{1}{2!}
\int d^2\xi _1
\int d^2\xi _2~
P(\boldsymbol{\theta }_1, \boldsymbol{\xi }_1)
Q(\boldsymbol{\xi }_1)
P(\boldsymbol{\theta }_2, \boldsymbol{\xi }_2)
Q(\boldsymbol{\xi }_2)\cr 
{\mathfrak p}_{2}&=&
\frac{1}{2!}
\int d^2\xi _1
P(\boldsymbol{\theta }_1, \boldsymbol{\xi }_1)
P(\boldsymbol{\theta }_2, \boldsymbol{\xi }_1)
Q(\boldsymbol{\xi }_1).
\ea 
Here ${\mathfrak p}_{1,1}(\boldsymbol{\theta }_1, \boldsymbol{\theta }_2)$
is the probability that the points $\boldsymbol{\theta }_1$ and
$\boldsymbol{\theta }_2$ are within different 
motifs centred at $\boldsymbol{\xi}_1$ and $\boldsymbol{\xi}_2,$ and 
${\mathfrak p}_{2}(\boldsymbol{\theta }_1, \boldsymbol{\theta }_2)$
is the probility that the points are within the same motif.
Analogously we define the three-point correlation function 
${\mathcal B}
(\boldsymbol{\theta}_1 ,\boldsymbol{\theta}_2 ,\boldsymbol{\theta}_3)$
by the sum of the three contributions 
\ba
{\mathcal B}
(\boldsymbol{\theta}_1 ,\boldsymbol{\theta}_2 ,\boldsymbol{\theta}_3)
={\mathfrak p}_{1,1,1}
(\boldsymbol{\theta}_1 ,\boldsymbol{\theta}_2 ,\boldsymbol{\theta}_3)
+{\mathfrak p}_{2,1}
(\boldsymbol{\theta}_1 ,\boldsymbol{\theta}_2 ,\boldsymbol{\theta}_3)
+{\mathfrak p}_3
(\boldsymbol{\theta}_1 ,\boldsymbol{\theta}_2 ,\boldsymbol{\theta}_3)
\ea
where
\ba
{\mathfrak p}_{1,1,1}&=&
\frac{1}{3!}
\int d^2\xi _1
\int d^2\xi _2~
\int d^2\xi _3~
P(\boldsymbol{\theta }_1, \boldsymbol{\xi }_1)
Q(\boldsymbol{\xi }_1)
P(\boldsymbol{\theta }_2, \boldsymbol{\xi }_2)
Q(\boldsymbol{\xi }_2)
P(\boldsymbol{\theta }_3, \boldsymbol{\xi }_3)
Q(\boldsymbol{\xi }_3)
\cr 
{\mathfrak p}_{2,1}&=&
\frac{1}{2!}
\frac{1}{2!}
\int d^2\xi _1
\int d^2\xi _2
P(\boldsymbol{\theta }_1, \boldsymbol{\xi }_1)
P(\boldsymbol{\theta }_2, \boldsymbol{\xi }_1)
Q(\boldsymbol{\xi }_1)
P(\boldsymbol{\theta }_3, \boldsymbol{\xi }_2)
Q(\boldsymbol{\xi }_2)
+ (\text{two permutations})\cr 
{\mathfrak p}_{3}&=&
\frac{1}{3!}
\int d^2\xi _1
P(\boldsymbol{\theta }_1, \boldsymbol{\xi }_1)
P(\boldsymbol{\theta }_2, \boldsymbol{\xi }_1)
P(\boldsymbol{\theta }_3, \boldsymbol{\xi }_1)
Q(\boldsymbol{\xi }_1)
\ea
are defined analogously.
The combinatorial factors preceding the integrals are the number of
equivalent motif configuration for indistinguishable patches in the
$\boldsymbol{\xi }$--space which must be introduced to avoid multiple
counting.
We introduce a  factor $1/n_{Q}!$ where $n_{Q}$ is the number of $Q$'s,
and a factor $1/n_{P}!$ for each $\boldsymbol{\xi }$ where $n_{P}$ is
the number of $P$'s shared by the same $\boldsymbol{\xi }.$
For convenience we also define the one-point function
\ba
\mathfrak{p}_{1}(\boldsymbol{\theta}_1)\equiv
\int d^2\xi _{1}~
 P(\boldsymbol{\theta }_1,\boldsymbol{\xi }_1) Q(\boldsymbol{\xi }_1).
\ea

\section{A Simple Model: Uniform Distribution}

For concreteness we assume the motifs to be circles of radius
$\theta_{circ},$
with sharp edges $P(\boldsymbol{\theta},\boldsymbol{\xi})
=N_{P}\Theta[\theta_{circ}-\vert\boldsymbol{\xi}-\boldsymbol{\theta }\vert]$ 
and uniformly distributed in the sky $Q(\boldsymbol{\xi})=const=N_{Q}.$ 
Thus $N_{Q}=\left<N\right>/A_{sky}$ and $N_{P}=T_{motif},$ which in
this case are constant inside the motif and independent of the
position of the motifs in the sky. [To be dimensionally correct, we
  must have $T_{motif}$ normalized to the average temperature of
  the sky $T_{sky}.$]
We will also discuss a non-uniform
distribution of the motifs in the sky, motivated by the variation of the
distribution of the large-scale structure along the latitude. This
will be the working case for introducing 
dependence of the temperature on the position of the motifs 
as encoded in the distribution probability. 
[For a refinement of the functional form see Ref.~\cite{bucher}.]

We compute the two and three-point correlator functions in real
space for the statistical ensemble just described.
The one-point function becomes
\ba
\mathfrak{p}_{1}(\boldsymbol{\theta}_1)
=N_{P}N_{Q}\int d^2x _1~
\Theta[\theta_{circ}-\vert\boldsymbol{x}_1\vert]
=N_{P}N_{Q}A_{motif}
\ea
where $A_{motif} =\pi\theta_{circ}^2.$
For the power spectrum we must calculate $\mathfrak{p}_{1,1}$ and
$\mathfrak{p}_2,$ which can be written as
\ba
\mathfrak{p}_{1,1}(\boldsymbol{\theta}_1,\boldsymbol{\theta}_2)&=&
{1\over 2!}N_{P}^2N_{Q}^2
\int d^2x _1~
\Theta[\theta_{circ}-\vert\boldsymbol{x}_1\vert]
\int d^2x _2~
\Theta[\theta_{circ}-\vert\boldsymbol{x}_2\vert]
={1\over 2!}\mathfrak{p}_{1}^2\
\\
\mathfrak{p}_{2}(\boldsymbol{\theta}_1,\boldsymbol{\theta}_2)&=&
\frac{1}{2!}N_{P}^2N_{Q}
\int d^2x _1~
\Theta[\theta_{circ}-\vert \boldsymbol{x}_1\vert]
\Theta[\theta_{circ}-\vert \boldsymbol{x}_1+\boldsymbol{\theta}_{12}\vert].
\ea 
Here $\boldsymbol{x}_i=\boldsymbol{\xi}_i -\boldsymbol{\theta} _i$ and
$\boldsymbol{\theta}_{ij}=\boldsymbol{\theta}_i-\boldsymbol{\theta}_j.$
For the bispectrum we observe that both $\mathfrak{p}_{1,1,1}$ and
$\mathfrak{p}_{2,1}$ can be expressed in terms of $\mathfrak{p}_{1}$ and
$\mathfrak{p}_{2}$ as follows 
\ba
\mathfrak{p}_{1,1,1}(\boldsymbol{\theta}_1,\boldsymbol{\theta} _2,
 \boldsymbol{\theta} _3)&=&
{1\over 3!}N_{P}^3N_{Q}^3
\int d^2x _1~
\Theta[\theta_{circ}-\vert\boldsymbol{x}_1\vert]
\int d^2x _2~
\Theta[\theta_{circ}-\vert\boldsymbol{x}_2\vert]\cr
&&\times\int d^2x _3~
\Theta[\theta_{circ}-\vert\boldsymbol{x}_3\vert]\cr
&=&{1\over 3!}\mathfrak{p}_{1}^3
\\
\mathfrak{p}_{2,1}(\boldsymbol{\theta}_1,\boldsymbol{\theta}_2,
 \boldsymbol{\theta} _3)&=&
\frac{1}{2!}\frac{1}{2!}N_{P}^3N_{Q}^2
\int d^2x _1~
\Theta[\theta_{circ}-\vert \boldsymbol{x} _1\vert]~
\Theta[\theta_{circ}-\vert \boldsymbol{x} _1+\boldsymbol{\theta} _{12}\vert]\cr
&&\times\int d^2x _3~
\Theta[\theta_{circ}-\vert\boldsymbol{x} _3\vert]
+ (\text{two permutations})\cr
&=&\frac{1}{2!}\mathfrak{p}_{1}\left[ 
\mathfrak{p}_{2}(\boldsymbol{\theta} _1,\boldsymbol{\theta} _2)
+\mathfrak{p}_{2}(\boldsymbol{\theta} _2,\boldsymbol{\theta} _3)
+\mathfrak{p}_{2}(\boldsymbol{\theta} _3,\boldsymbol{\theta} _1)
\right],
\ea
so we have in addition to compute $\mathfrak{p}_{3}$ only
\ba
\mathfrak{p}_{3}(\boldsymbol{\theta} _1,\boldsymbol{\theta} _2,
 \boldsymbol{\theta} _3)=
\frac{1}{3!}N_{P}^3N_{Q}
\int d^2x _1
\Theta[\theta_{circ}-\vert \boldsymbol{x} _1\vert]~
\Theta[\theta_{circ}-\vert \boldsymbol{x} _1+\boldsymbol{\theta} _{12}\vert]~
\Theta[\theta_{circ}-\vert \boldsymbol{x}_1+\boldsymbol{\theta} _{13}\vert].
\ea
These quantities describe the probability that one, two or three points, 
denoted by $\boldsymbol{\theta} _1,$ $\boldsymbol{\theta} _2$ and
$\boldsymbol{\theta} _3,$ are within the same motif
centred at points $\boldsymbol{\xi}$'s which are 
distributed in the sky according to $Q.$  
These probabilities can also be interpreted as the measure of the overlap of
motifs centred at each of these points $\boldsymbol{\theta}$'s.
The integrals of the step functions about the distance from the
centres at $\boldsymbol{\xi}$'s 
can thus be visualized as the area of the intersection of motifs centred
at the $\boldsymbol{\theta }$'s. [For more details see Ref.~\cite{bucher}.]
We find that
\ba
\mathfrak{p}_{1,1}(\boldsymbol{\theta} _1,\boldsymbol{\theta} _2)&=&
{1\over 2!}\mathfrak{p}_{1}^2\\
\mathfrak{p}_{2}(\boldsymbol{\theta} _1,\boldsymbol{\theta} _2)&=&
{1\over 2!}
N_{P}^2N_{Q}A_{1\cap 2}
\ea
where $A_{1\cap 2}$ is the area of intersection  between the motifs
 centred at $\boldsymbol{\theta} _{1}$ and $\boldsymbol{\theta} _{2}$
\ba
A_{1\cap 2}
=\left( 2\theta_{circ}^2
  \arctan\left[ \sqrt{ {4\theta_{circ}^2\over 
   {\vert\boldsymbol{\theta} _{12}\vert ^2}} -1}\right] 
-{{\vert\boldsymbol{\theta} _{12}\vert ^2}\over 2}
  \sqrt{ {4\theta_{circ}^2\over 
   {\vert\boldsymbol{\theta} _{12}\vert ^2}} -1}\right)
 \Theta\left[ 2\theta _{circ} -\vert\boldsymbol{\theta} _{12}\vert\right].
\ea
These contribute to the power spectrum. Moreover, we find that
\ba
\mathfrak{p}_{1,1,1}(\boldsymbol{\theta} _1,\boldsymbol{\theta} _2,
 \boldsymbol{\theta} _3)&=&
{1\over 3!}\mathfrak{p}_{1}^3\\
\mathfrak{p}_{2,1}(\boldsymbol{\theta} _1,\boldsymbol{\theta} _2,
 \boldsymbol{\theta} _3)&=&
{1\over 2!}\mathfrak{p}_{1}
 \left[
  \mathfrak{p}_{2}(\boldsymbol{\theta} _1,\boldsymbol{\theta} _2)
  +\mathfrak{p}_{2}(\boldsymbol{\theta} _2,\boldsymbol{\theta} _3)
  +\mathfrak{p}_{2}(\boldsymbol{\theta} _3,\boldsymbol{\theta} _1)
 \right]\\
\mathfrak{p}_{3}(\boldsymbol{\theta} _1,\boldsymbol{\theta} _2,
 \boldsymbol{\theta} _3)&=&{1\over 3!}N_{P}^3N_{Q}A_{1\cap 2\cap 3}
\ea
where $A_{1\cap 2\cap 3}$ is the area of intersection among the three motifs
\ba
A_{1\cap 2\cap 3}
&=&\left[ {1\over 4}
 \sqrt{\left( 
 \vert\boldsymbol{\theta} _{12}\vert ^2
 +\vert\boldsymbol{\theta} _{23}\vert ^2
 +\vert\boldsymbol{\theta} _{13}\vert ^2\right)^2
 -2\left(
 \vert\boldsymbol{\theta} _{12}\vert ^4
 +\vert\boldsymbol{\theta} _{23}\vert ^4
 +\vert\boldsymbol{\theta} _{13}\vert ^4\right)}
-{1\over 2}A_{motif}\right]\cr
&&\times
\Theta\left[ 2\theta _{circ} -\vert\boldsymbol{\theta} _{12}\vert\right]
 \Theta\left[ 2\theta _{circ} -\vert\boldsymbol{\theta} _{23}\vert\right]
  \Theta\left[ 2\theta _{circ} -\vert\boldsymbol{\theta} _{13}\vert\right]\cr
&+&{1\over 2}A_{1\cap 2}
 \Theta\left[ 2\theta _{circ} -\vert\boldsymbol{\theta} _{23}\vert\right]
  \Theta\left[ 2\theta _{circ} -\vert\boldsymbol{\theta} _{13}\vert\right]
\cr 
&+&{1\over 2}A_{2\cap 3}
 \Theta\left[ 2\theta _{circ} -\vert\boldsymbol{\theta} _{12}\vert\right]
  \Theta\left[ 2\theta _{circ} -\vert\boldsymbol{\theta} _{13}\vert\right]
\cr 
&+&{1\over2}A_{1\cap 3}
 \Theta\left[ 2\theta _{circ} -\vert\boldsymbol{\theta} _{12}\vert\right]
 \Theta\left[ 2\theta _{circ} -\vert\boldsymbol{\theta} _{23}\vert\right].
\ea
These contribute to the bispectrum.
It follows that
\ba
\mathcal{P}(\boldsymbol{\theta} _1,\boldsymbol{\theta} _2)&=&
{1\over 2!}\mathfrak{p}_{1}^2
+\mathfrak{p}_2(\boldsymbol{\theta} _1,\boldsymbol{\theta} _2)
\\
\mathcal{B}(\boldsymbol{\theta} _1,\boldsymbol{\theta} _2,
\boldsymbol{\theta} _3)&=&
{1\over 3!}\mathfrak{p}_{1}^3
+{1\over 2!}\mathfrak{p}_1\left[
\mathfrak{p}_2(\boldsymbol{\theta} _1,\boldsymbol{\theta} _2)
+\mathfrak{p}_2(\boldsymbol{\theta} _2,\boldsymbol{\theta} _3)
+\mathfrak{p}_2(\boldsymbol{\theta}_3,\boldsymbol{\theta} _1)\right]
+\mathfrak{p}_3(\boldsymbol{\theta}_1,\boldsymbol{\theta} _2,
\boldsymbol{\theta} _3).
\ea
 
We now proceed to compute the value of $f_{NL}$ produced. 
The parameter $f_{NL}$ characterizes the amplitude of the
temperature non-Gaussianity since it couples to the quadratic term of
the expansion of the temperature fluctuation 
$\mathcal{T}\equiv \Delta T/T$ about a Gaussian distribution
$\mathcal{T}_{L}.$  
Assuming the Sachs--Wolfe approximation $(\Delta T)/T =-(1/3)\Delta
\Phi/c^2$ 
on all scales 
and for an infinitely thin surface of last
scattering, we have in real space that
\ba
\mathcal{T}(\boldsymbol{\theta})
=\mathcal{T}_{L}(\boldsymbol{\theta})
+3f_{NL}\left[ \mathcal{T}_{L}^2(\boldsymbol{\theta}) 
-\left< \mathcal{T}_{L}(\boldsymbol{\theta})\right>^2\right].
\ea
Here the Gaussian distribution has zero mean, 
$\left< \mathcal{T}_{L}(\boldsymbol{\theta})\right>=0,$ from which it
follows to leading order that 
\ba
\mathcal{P}(\boldsymbol{\theta} _1,\boldsymbol{\theta} _2)\equiv 
\left< \mathcal{T}(\boldsymbol{\theta} _1)
 \mathcal{T}(\boldsymbol{\theta} _2)\right>
=\left< \mathcal{T}_{L}(\boldsymbol{\theta} _1)
  \mathcal{T}_{L}(\boldsymbol{\theta} _2 )\right>.
\ea
Since a Gaussian distribution has vanishing odd-order momenta, 
we find for the three-point correlation function that
\ba
&&\mathcal{B}(\boldsymbol{\theta} _1,\boldsymbol{\theta} _2,
\boldsymbol{\theta} _3)\equiv
\left< \mathcal{T}(\boldsymbol{\theta} _1)
 \mathcal{T}(\boldsymbol{\theta} _2)
  \mathcal{T}(\boldsymbol{\theta} _3 )\right>\cr
&=&6f_{NL}\left[ 
\mathcal{P}(\boldsymbol{\theta} _1,\boldsymbol{\theta} _3)
 \mathcal{P}(\boldsymbol{\theta} _3,\boldsymbol{\theta} _2)
+\mathcal{P}(\boldsymbol{\theta} _2,\boldsymbol{\theta} _1)
  \mathcal{P}(\boldsymbol{\theta} _1,\boldsymbol{\theta} _3)
+\mathcal{P}(\boldsymbol{\theta} _3,\boldsymbol{\theta} _2)
  \mathcal{P}(\boldsymbol{\theta} _2,\boldsymbol{\theta} _1)\right].
\ea
We distinguish three cases for the three possible relations among the
distances between the points $\theta's,$ namely  
1) $\theta_{12},\theta_{13}, \theta_{23}> 2\theta_{circ};$
2) $\theta_{12}< 2\theta_{circ}$ 
and $\theta_{13},\theta_{23}> 2\theta_{circ};$
3) $\theta_{12},\theta_{13},\theta_{23}< 2\theta_{circ}.$
Let us fix the size of the motif and for concreteness take the radius to
be of the size of the resolution of the map, i.e. $\theta_{circ}\sim
10^{-6}.$

\subsection{Case $\theta_{12},\theta_{13},\theta_{23}> 2\theta_{circ}$}

This is the case where each point $\boldsymbol{\theta}_i$ is within a
different motif and consequently only the one-point function 
$\mathfrak{p}_{1}$ and the contributions to the two and three-point
functions which can be expressed in terms of $\mathfrak{p}_{1},$ i.e.
$\mathfrak{p}_{1,1}$ and $\mathfrak{p}_{1,1,1},$
do not vanish. We find that
\ba
\mathcal{P}(\boldsymbol{\theta} _1,\boldsymbol{\theta} _2)
&=&\mathcal{P}(\boldsymbol{\theta} _1,\boldsymbol{\theta} _3)
=\mathcal{P}(\boldsymbol{\theta} _2,\boldsymbol{\theta} _3)
= \mathfrak{p}_{1,1} ={1\over 2}\mathfrak{p}_{1}^2, \cr
\mathcal{B}(\boldsymbol{\theta} _1,\boldsymbol{\theta} _2,
 \boldsymbol{\theta} _3)
&=& \mathfrak{p}_{1,1,1} ={1\over 6}\mathfrak{p}_{1}^3,
\ea
from which it follows that
\ba
f_{NL}
={1\over 6}
 {\mathcal{B}(\boldsymbol{\theta} _1,\boldsymbol{\theta} _2, 
  \boldsymbol{\theta} _3) \over 
  {3\mathcal{P}(\boldsymbol{\theta} _1,\boldsymbol{\theta} _2)
     \mathcal{P}(\boldsymbol{\theta} _2,\boldsymbol{\theta} _3)}}
={1\over 27}{1\over \mathfrak{p}_{1}},
\quad
\mathfrak{p}_{1}> 1.
\ea


\subsection{Case $\theta_{12}< 2\theta_{circ}$ 
and $\theta_{13},\theta_{23}> 2\theta_{circ}$}

This is the case where two points, here for concreteness
$\boldsymbol{\theta} _{1}$ and $\boldsymbol{\theta} _{2},$ are within
the same motif. In addition to the one-point function and the
contributions derived from it, we have 
$\mathfrak{p}_{2}(\boldsymbol{\theta} _1,\boldsymbol{\theta} _2)\not=0$ 
and
$\mathfrak{p}_{2,1}(\boldsymbol{\theta} _1,\boldsymbol{\theta} _2,
 \boldsymbol{\theta} _3)\not=0$ from the contribution to the
 three-point function of the two-point function between 
$\boldsymbol{\theta} _{1}$ and $\boldsymbol{\theta} _{2}$ only. 
Hence
\ba
\mathcal{P}(\boldsymbol{\theta} _1,\boldsymbol{\theta} _2)
&=& \mathfrak{p}_{1,1} 
+\mathfrak{p}_{2}(\boldsymbol{\theta} _1,\boldsymbol{\theta} _2)
= {1\over 2}\mathfrak{p}_{1}^2 +\mathfrak{p}_{2},\quad
\mathcal{P}(\boldsymbol{\theta} _1,\boldsymbol{\theta} _3)
=\mathcal{P}(\boldsymbol{\theta} _2,\boldsymbol{\theta} _3)
= \mathfrak{p}_{1,1}
={1\over 2}\mathfrak{p}_{1}^2, \cr
\mathcal{B}(\boldsymbol{\theta} _1,\boldsymbol{\theta} _2,
 \boldsymbol{\theta} _3)
&=& \mathfrak{p}_{1,1,1} 
+ \mathfrak{p}_{2,1} (\boldsymbol{\theta} _1,\boldsymbol{\theta} _2,
 \boldsymbol{\theta} _3)
={1\over 6}\mathfrak{p}_{1}^3
+{1\over 2}\mathfrak{p}_{1}
  \mathfrak{p}_{2}(\boldsymbol{\theta} _1,\boldsymbol{\theta} _2),
\ea
and
\ba
f_{NL} 
= {1\over 6}
{\mathcal{B}(\boldsymbol{\theta} _1,\boldsymbol{\theta} _2,
 \boldsymbol{\theta} _3)\over
 {2\mathcal{P}(\boldsymbol{\theta} _1,\boldsymbol{\theta} _2)
    \mathcal{P}(\boldsymbol{\theta} _2,\boldsymbol{\theta} _3)
  +\mathcal{P}(\boldsymbol{\theta} _1,\boldsymbol{\theta} _3)
    \mathcal{P}(\boldsymbol{\theta} _2,\boldsymbol{\theta} _3)}}
={1\over 6}{
{ {1\over 6}\mathfrak{p}_{1}^3 
 +{1\over 2}\mathfrak{p}_{1}
  \mathfrak{p}_{2}(\boldsymbol{\theta} _1,\boldsymbol{\theta} _2)
}
   \over 
{ 2\left( {1\over 2}\mathfrak{p}_{1}^2
 +\mathfrak{p}_{2}(\boldsymbol{\theta} _1,\boldsymbol{\theta} _2)
\right){1\over 2}\mathfrak{p}_{1}^2 
+{1\over 4}\mathfrak{p}_{1}^4}
}.
\ea

We note that  
$\mathfrak{p}_{1}^2\propto N_{Q}^2A_{motif}^2$ and 
$\mathfrak{p}_{2}\propto N_{Q}A_{1\cap 2}<N_{Q}A_{motif},$ where
$N_{Q}$ is the mean number density of motifs in the sky.
The relative magnitude of $\mathfrak{p}_{1}^2$ and $\mathfrak{p}_{2}$
depends on the relation between the mean number of motifs 
$\left< N\right>$ and the area of the motif $A_{motif}.$ 
For convenience we define 
$\alpha\equiv N_{Q}A_{motif}=\left< N\right>A_{motif}/A_{sky}.$
We must further distinguish between the following two cases.
If $\alpha <\alpha^2,$ i.e. $\left< N\right> >A_{sky}/A_{motif} \sim
10^{13},$ then $\mathfrak{p}_{2}<\alpha<\mathfrak{p}_{1}^2.$
This suggests that a minimum density of motifs is required in order
for the one-point function to dominate, i.e. in order for two points
which are close enough to be within the same motif to be also within
a second, and thus necessarily overlapping, motif.
On the other hand if $\alpha >\alpha^2,$ i.e. $\left< N\right>
<A_{sky}/A_{motif} \sim 10^{13},$ then both
$\mathfrak{p}_{1}^2,\mathfrak{p}_{2}<\alpha.$ Here in order to discriminate
the relative magnitude, we need in addition to precise the relation
between $A_{1\cap2}$ and $A_{motif},$ which depends on
$\theta_{12}.$ If $\theta_{12}$ is sufficiently small so that
  $A_{1\cap2}\sim A_{motif},$ then $\mathfrak{p}_{2}\sim \alpha$ and
consequently $\mathfrak{p}_{2}>\mathfrak{p}_{1}^2.$ However, if
$\theta_{12}$ is very close to $2\theta_{circ},$  then 
$A_{1\cap2}\ll A_{motif}$ and consequently 
$\mathfrak{p}_{2}\ll \alpha,$ so that if $\mathfrak{p}_{2}<\alpha^2$ 
then $\mathfrak{p}_{2}<\mathfrak{p}_{1}^2.$ This is equivalent
to the case  $\alpha <\alpha^2.$

Hence, if $\left< N\right> >10^{13},$ or $\left< N\right> <10^{13}$ and
$\theta_{12}\sim 2\theta_{circ},$ then
\ba
f_{NL}
\sim{1\over 6}
{ { {1\over 6}\mathfrak{p}_{1}^3} 
 \over {{3\over 4}\mathfrak{p}_{1}^4}}
={1\over 27}{1\over \mathfrak{p}_{1}}, 
\quad
\mathfrak{p}_{1}> 1,
\ea
whereas if $\left< N\right> <10^{13}$ and $\theta_{12}\ll 2\theta_{circ},$ 
then
\ba
f_{NL}
\sim {1\over 6}
{ { {1\over 2}\mathfrak{p}_{1}
\mathfrak{p}_{2}(\boldsymbol{\theta} _{1},\boldsymbol{\theta}_{2})}   
 \over {\mathfrak{p}_{2}(\boldsymbol{\theta} _{1},\boldsymbol{\theta}_{2})
   \mathfrak{p}_{1}^2}}
={1\over 12}{1 \over \mathfrak{p}_{1}},
\quad
\mathfrak{p}_{1}< 1.
\ea


\subsection{Case $\theta_{12},\theta_{13},\theta_{23}< 2\theta_{circ}$}

This is the case where the three points are within the same
motif. All the terms contribute to the two and three-point functions.
In order to determine which contribution dominates for a mean number
of motifs $\left< N\right>$ distributed in the sky according to 
$Q,$ we do an analysis similar to that above. 
Thus if  $\left< N\right> >10^{13}$
\ba
f_{NL}
\sim {1\over 6}
{ {1\over 6}\mathfrak{p}_{1}^3
 \over { {3\over 4}\mathfrak{p}_{1}^4}
}
={1\over 27}{1\over \mathfrak{p}_{1}},
\quad
\mathfrak{p}_{1}> 1.
\ea
However, if $\left< N\right> <10^{13}$ we find 
for $\theta _{12}<\theta _{13}, \theta _{23}$ that
\ba
f_{NL}
\sim {1\over 6}
{ \mathfrak{p}_{3}(\boldsymbol{\theta} _{1},\boldsymbol{\theta} _{2},
  \boldsymbol{\theta} _{3})
 \over 
 {2\mathfrak{p}_{2}(\boldsymbol{\theta} _{1},\boldsymbol{\theta} _{2})
  \mathfrak{p}_{2}(\boldsymbol{\theta} _{2},\boldsymbol{\theta} _{3})}
}
={1\over 18}
{ { -{1\over 2}A_{motif} +{3\over 2}A_{1\cap2}}
 \over {N_{P}N_{Q}A_{1\cap2}A_{2\cap3}}},
\ea
while for $\theta _{12}\sim \theta _{23}\sim \theta _{13}$ we find that 
\ba
f_{NL}
\sim {1\over 6}
{ \mathfrak{p}_{3}(\boldsymbol{\theta} _{1},\boldsymbol{\theta} _{2},
  \boldsymbol{\theta} _{3})
 \over 
 {3\mathfrak{p}_{2}(\boldsymbol{\theta} _{1},\boldsymbol{\theta} _{2})
  \mathfrak{p}_{2}(\boldsymbol{\theta} _{2},\boldsymbol{\theta} _{3})}
}
={1\over 27}{ 
{ {\sqrt{3}\over 4} \theta _{12}^2 -{1\over 2}A_{motif}+{3\over 2}A_{1\cap2}}
 \over {N_{P}N_{Q}A_{1\cap 2}^2}}.
\ea
For small $\theta_{12}$ the term $(\sqrt{3/4})\theta_{12}^2$ is
subdominant, so for both cases we need only compare $A_{1\cap 2}$ with
$A_{motif}.$  We present the results for the former case, with those for the
latter following straighfowardly.
For $\theta<\theta_{eq},$ where $\theta_{eq}$ is the distance such
that $(3/2)A_{1\cap2}=(1/2)A_{motif},$ we find that
$(3/2)A_{1\cap2}>(1/2)A_{motif}$ and hence that 
\ba
f_{NL}\sim{1\over 18}
{ {A_{1\cap2}}
 \over {N_{P}N_{Q}A_{1\cap2}A_{2\cap3}}}
={1\over 18}
{ {1}
 \over {N_{P}N_{Q}A_{2\cap3}}}.
\ea
For $\theta_{eq}<\theta_{12}<2\theta_{circ}$ the contribution of
$(\sqrt{3/4})\theta_{12}^2$ becomes more important but the dominant
contribution is that of $(1/2)A_{motif}.$ Moreover since $A_{1\cap2}$
is a decreasing function of $\theta_{12}$ except for $\theta_{12}$
very close to $2\theta_{circ},$ then
$A_{1\cap2}A_{2\cap3}>A_{2\cap3}^2$ and consequently 
$-1/A_{1\cap2}A_{2\cap3}>-1/A_{2\cap3}^2.$  Thus 
\ba
f_{NL}\gtorder {1\over 18}
{ {-{1\over 2}A_{motif}}
 \over {N_{P}N_{Q}A_{1\cap2}A_{2\cap3}}}
>{1\over 18}
{ {-{1\over 2}A_{motif}}
 \over {N_{P}N_{Q}A_{2\cap3}^2}}.
\ea
A similar calculation would follow for the case
$\theta _{12}\sim \theta _{23}\sim \theta _{13}.$\\


Since these distances must be such that 
$0\le\theta _{12},\theta _{13},\theta _{23}\le 2\theta_{circ},$ then 
$-\infty\le f_{NL}<\infty,$ i.e. we can generate arbitrary
non-Gaussianity. We need, however, to be able to constrain 
$f_{NL}$ to a finite interval.
This requires the use of the foreground maps to constrain the
parameters for the case of the uniform model, as well as
the functional form of the probability functions for more realistic
models.

\section{Constraining the Model}

Constraints on the parameters of the model can be extracted from the
properties of foreground maps. 
Spatial templates from the WMAP data 
produced for synchroton, free-free and dust emission 
show a temperature distribution strongly dependent on the latitute. 
In order to account for this observation we must consider a
non-uniform probability density that can capture the qualitative
aspects of these dependence.
The simplest case of a non-uniform distribution of motifs
in the sky is that where $Q(\boldsymbol{\theta})$ is a slowly varying function
acroos the scale of the motif, so that the
probabilities are changed to 
\ba
\mathfrak{p}_{1,1}(\boldsymbol{\theta} _1,\boldsymbol{\theta} _2)&=&
{1\over 2!}N_{P}^2A_{motif}^2
 Q(\boldsymbol{\theta} _1)Q(\boldsymbol{\theta} _2)\\
\mathfrak{p}_{2}(\boldsymbol{\theta} _1,\boldsymbol{\theta} _2)&=&
{1\over 2!}N_{P}^2A_{1\cap 2}
Q\left( { {\boldsymbol{\theta} _1 +\boldsymbol{\theta} _2}\over 2}\right)
\ea
and to
\ba
\mathfrak{p}_{1,1,1}(\boldsymbol{\theta} _1,\boldsymbol{\theta} _2,
 \boldsymbol{\theta} _3)&=&
{1\over 3!}N_{P}^3A_{motif}^3
 Q(\boldsymbol{\theta} _1)Q(\boldsymbol{\theta} _2)
  Q(\boldsymbol{\theta} _3)\\
\mathfrak{p}_{2,1}(\boldsymbol{\theta} _1,\boldsymbol{\theta} _2,
 \boldsymbol{\theta} _3)&=&
{1\over 2!}N_{P}A_{motif}
\cr
&&\times\biggl[
  \mathfrak{p}_{2}(\boldsymbol{\theta} _1,\boldsymbol{\theta} _2)
    Q(\boldsymbol{\theta} _3)
  +\mathfrak{p}_{2}(\boldsymbol{\theta} _2,\boldsymbol{\theta} _3)
      Q(\boldsymbol{\theta} _1)
  +\mathfrak{p}_{2}(\boldsymbol{\theta} _3,\boldsymbol{\theta} _1)
    Q(\boldsymbol{\theta} _2)
 \biggr]\\
\mathfrak{p}_{3}(\boldsymbol{\theta} _1,\boldsymbol{\theta} _2,
 \boldsymbol{\theta} _3)&=&
{1\over 3!}N_{P}^3A_{1\cap 2\cap 3}Q\left( {{\boldsymbol{\theta} _1
+\boldsymbol{\theta} _2 
 +\boldsymbol{\theta} _3}\over 3}\right).
\ea
The changes in the two and three-point correlation functions, as well
as in $f_{NL},$ are straightfoward.
Assuming that the contribution of the overlapping motifs add in the
intersection region (what we called the ``optically thin'' approximation),
then the value of $Q$ in the mid distance between the centres of two
overlapping motifs is the sum of the value of $Q$ at the centre of each
motif, i.e.
\ba
Q\left( {{\boldsymbol{\theta} _i +\boldsymbol{\theta} _j}\over 2}\right)
=Q\left( \boldsymbol{\theta} _i\right) 
+Q\left( \boldsymbol{\theta} _j\right).
\ea
Since the scale on which $Q\left(\boldsymbol{\xi}\right)$ varies is
large compared to $A_{motif},$ then
\ba
Q(\boldsymbol{\xi})= 
{{\left< T\left(\boldsymbol{\xi}\right)\right>} \over T_{motif}}
{1\over A_{sky}}
\ea
so that the same normalization condition holds.
Possible functional forms for  $Q\left(\boldsymbol{\xi}\right)$
will be explored in Ref.~\cite{bucher}. For the purpose of this
section, however, we will not need to be more specific.
The conditional probability becomes
$P(\boldsymbol{\theta},\boldsymbol{\xi})=
T(\boldsymbol{\theta})
 \Theta\left[\theta_{circ}
  -\vert\boldsymbol{\xi}-\boldsymbol{\theta }\vert\right]$ 
which will be simply the average temperature at
the point $\boldsymbol{\xi}$ where the motif on which
$\boldsymbol{\theta}$ lies is centred.

By quantifying the clean regions of these maps, we can also set bounds
on the foreground temperature.
From a carefull examination of the five-year data maps \cite{wmap5:gold} 
we extracted
the minimum temperature $T_{0}$ that 
a foreground source 
should have in order to saturate the map with as close to a single
temperature value as possible.  
Taking the minimum value of the
temperature of the three foregrounds for each of the five frequency
bands analysed, we found by eye inspection that saturation of the maps
was achieved for $T_{0}$ a few $\mu K.$
Hence the temperature at each point of the motif distribution must be
such that $T( \boldsymbol{\xi})>T_{0}.$
On the other hand, the temperature of a single spot must be highly constrained
so as not to stand out. 
This observation is captured by the condition that the power
spectrum generated by the motif distribution does not outshine the
background radiation. If the motif distribution produced a large power
spectrum, as that produced by small and bright sources, then non-Gaussianity
would be obvious. We thus set
$\mathcal{P}( \boldsymbol{\theta}_{i},\boldsymbol{\theta}_{j}) 
<(T_{0}/T_{sky})^2.$

The three cases discriminated in the previous section will now be
combined with the constraints
on the temperature obtained from the foreground maps. 
Apart from factors of order $O(1),$ the results will be the same for
both a uniform and a non-uniform probability distribution of motifs in
the sky. For simplicity we will analyse the results from the uniform
distribution.

\subsection{Case $\theta_{12},\theta_{13},\theta_{23}> 2\theta_{circ}$}

Here for any two pair of motif centres $i,j,$ the constraints on the
temperature yield that
\ba
\mathcal{P}( \boldsymbol{\theta}_{i},\boldsymbol{\theta}_{j}) 
={1\over 2}N_{P}^2N_{Q}^2A_{motif}^2
>{1\over 2}\left( {T_{0}\over T_{sky}}\right)^2A_{motif}^2N_{Q}^2
\ea
and simultaneously that
\ba
\mathcal{P}( \boldsymbol{\theta}_{i},\boldsymbol{\theta}_{j}) 
<\left( {T_{0}\over T_{sky}}\right)^2
\label{eq:powerconstraint}
\ea
from which it follows that
\ba
{1\over 2}\left< N\right>^2
 \left( {A_{motif}\over A_{sky}}\right)^2
<1.
\label{eq:powerconstraint1}
\ea
This relation constrains the size of the motifs given their average
number $\left< N\right>$ distributed on the sky.
Using the relations above in the expression for $f_{NL}$ we find that
\ba
f_{NL}={1\over 27}{1\over {N_PA_{motif}N_Q
}}
<{1\over 27}{1\over {\left< N\right>
 (A_{motif}/A_{sky})(T_{0}/T_{sky})}}.
\ea
Given the constraint from the power spectrum, we find that the
magnitude of $f_{NL}$ is further determined by $(T_{0}/T_{sky})$ only. Here
$f_{NL}$ is thus a measure of the deviation of $T_{0}$ from the average sky
temperature.

\subsection{Case $\theta_{12}< 2\theta_{circ}$ 
and $\theta_{13},\theta_{23}> 2\theta_{circ}$}

Here for $\left< N\right> >10^{13},$
or $\left< N\right> <10^{13}$ and $\theta _{12}\sim 2\theta_{circ}$
\ba
\mathcal{P}( \boldsymbol{\theta}_{1},\boldsymbol{\theta}_{2}) 
\sim \mathcal{P}( \boldsymbol{\theta}_{1},\boldsymbol{\theta}_{3}) 
=\mathcal{P}( \boldsymbol{\theta}_{2},\boldsymbol{\theta}_{3}) 
={1\over 2}N_{P}^2A_{motif}^2N_{Q}^2
>{1\over 2}\left( {T_{0}\over T_{sky}}\right)^2A_{motif}^2N_{Q}^2,
\ea
which reduces to the previous case.
For 
$\left< N\right> <10^{13}$ and $\theta _{12}\ll\theta_{circ},$ however,
\ba
\mathcal{P}( \boldsymbol{\theta}_{1},\boldsymbol{\theta}_{2}) 
\sim {1\over 2}N_{P}^2A_{1\cap2}N_{Q}
>\left( {T_{0}\over T_{sky}}\right)^2A_{1\cap2}N_{Q}
\ea
which together with Eqn.~(\ref{eq:powerconstraint})
yields
\ba
{1\over 2}\left< N\right>{A_{1\cap2}\over A_{sky}}<1
\label{eq:powerconstraint2}
\ea
This relation constrains the overlapping between any two motifs 
given the average number of motifs.
Then for $f_{NL}$ we find that
\ba
f_{NL}\sim {1\over 12}{1\over {N_{P}A_{motif}N_{Q}
}}
<{1\over 12}{1\over {\left< N\right>
 (A_{motif}/A_{sky})(T_{0}/T_{sky})}}.
\label{eq:fNL2.2}
\ea
This relation sets a constraint on $f_{NL}$ of the same order of
magnitude as the previous case. 

\subsection{Case $\theta_{12},\theta_{13},\theta_{23}< 2\theta_{circ}$}

Here for $\left< N\right> >10^{13}$ we find the same result as in the
first case where the constraint on the power spectrum is given by 
Eqn.~(\ref{eq:powerconstraint1}).
For  $\left< N\right> <10^{13}$ the constraint on the power spectrum
is the same as in the second example of the second case and given by 
Eqn.~(\ref{eq:powerconstraint2}). We further discriminate between two
cases, namely $\theta _{12}<\theta _{13}, \theta _{23}$ and
$\theta _{12}\sim \theta _{13}\sim \theta _{23},$ and each case for
two regimes of the parameter $\theta_{12}.$ Thus for 
$\theta _{12}<\theta _{13}, \theta _{23}$ we find that
\ba
f_{NL}&\sim& {1\over 18}
 {{ A_{1\cap2\cap3}
  }\over {N_{P}A_{1\cap2}
    A_{2\cap3}
    N_{Q}
 }}
<{1\over 18}
 { {-{1\over 2}A_{motif}+{3\over 2}A_{1\cap2}}
 \over {\left< N\right>A_{1\cap2}(A_{2\cap3}/A_{sky})(T_{0}/T_{sky})}
 },
\ea
which for $\theta_{12}<\theta_{eq}$ becomes
\ba
f_{NL} <{1\over 18}
 { 1
 \over {\left< N\right>(A_{2\cap3}/A_{sky})(T_{0}/T_{sky})}
 },
\label{eq:fNL3.1.1}
\ea
whereas for $\theta_{eq}<\theta_{12}<2\theta_{circ}$
\ba
f_{NL}\gtorder {1\over 18}
 { {-{1\over 2}A_{motif}}
 \over {\left< N\right>A_{1\cap2}(A_{2\cap3}/A_{sky})(T_{0}/T_{sky})}
 }
>  {1\over 18}
 { {-{1\over 2}A_{motif}}
 \over {\left< N\right>A_{2\cap3}(A_{2\cap3}/A_{sky})(T_{0}/T_{sky})}
 }.
\label{eq:fNL3.1.2}
\ea
A similar calculation would follow straightfowardly for the case
$\theta _{12}\sim \theta _{13}\sim \theta _{23}.$ 
Eqn.~(\ref{eq:fNL3.1.1}) sets a weaker constraint
on $f_{NL}$ than Eqn.~(\ref{eq:fNL2.2}) by the order of magnitude of
$A_{2\cap3}/A_{motif}$.
Since $A_{1\cap 2}$ is predominantly a decreasing function of
$\theta_{12},$ Eqn.~(\ref{eq:fNL3.1.2}) sets a weaker still constrain
while allowing for a negative correlation among the three motifs.


\section{Discussion}

We propose a simple family of models for mimicking foreground sources that could
contaminate the non-Gaussian signal of the CMB. This contamination
could lead to the misidentification of a detection of a non-Gaussian
signal for primordial when in fact we would be looking at the spurious
signal from late-time, non-linear sources. 
Qualitatively non-gaussian aspects of foregrounds that are likely to
give a significant signal of $f_{NL}\not=0$ result from the modulation
of the small-scale power by the large-scale power. 

Our model allows to generate foreground maps by distributing motifs in
the sky according to a probability density  $Q(\boldsymbol{\xi})$ and
correlating them according to a conditional probability
$P(\boldsymbol{\theta,\xi})$ of finding one, two or three
points inside the same motif. The statistical properties of the
resulting maps are determined by the two and three-point correlation
functions, which are calculated by simply evaluated tree-level Feynman
diagrams.
We find the expression for $f_{NL}$ in terms of the parameters of 
$Q(\boldsymbol{\xi})$ and $P(\boldsymbol{\theta,\xi}),$ namely the
mean number of motifs and the intersection areas among two and three
motifs. We suggest a
prescription for introducing temperature in order to interpret the
statistical properties of the motif ensemble as statistical properties
of the temperature anisotropies. We also indicate how to use the
foreground maps to constrain the normalization factors and accordingly
we constrained the values for $f_{NL}.$ 

In the forthcoming paper
we will use the model to generate concrete mock foregroun maps
consistent with the level of foreground contamination observed and
make detailed analysis of the impact on $f_{NL}$ for {\it Planck} and
other experiments.









\ack

The author thanks Martin Bucher and Bartjan van Tent for useful discussions
and insightful comments.
The author was supported
by the Funda\c c\~ao para a Ci\^encia e a Tecnologia 
under the fellowship /BPD/18236/2004.
The author also acknowledges the use of the Legacy Archive for
Microwave Backround Data Analysis (LAMBDA) and the HEALPix package
\cite{healpix}.

\end{document}